\documentclass[runningheads]{svmult}
\usepackage{makeidx}   %
\usepackage{graphicx}  %
\usepackage{subeqnar}  %
\usepackage{physprbb}  %
\usepackage{times}
\usepackage{helvetic}

\def\yy#1{\relax}
\def\yyyy#1{\relax}
\def\yyyy#1{{\bf[[#1]]}}
\def\yyyyyy#1{\relax}
\def\yyyyyy#1{{\bf[[#1]]}}

\def\mede#1{\left\langle#1\right\rangle}

\begin{document}
\title*{%
Breakdown and recovery in traffic flow models\thanks{To appear in:
Traffic and Granular Flow '01, edited by Y. Sugiyama et al}
}

\author{%
Kai Nagel\inst{1}\thanks{nagel@inf.ethz.ch}
\and Christopher Kayatz\inst{1}
\and Peter Wagner\inst{2}\thanks{peter.wagner@dlr.de}
}
\institute{%
Dept.\ of Computer Science, ETH Z\"urich, 8092 Z\"urich, Switzerland
\and Institute for Transportation Research, German Aerospace Centre,
12489 Berlin, Germany}

\maketitle

\begin{abstract}
  Most car-following models show a transition from laminar to
  ``congested'' flow and vice versa.  Deterministic models often have
  a density range where a disturbance needs a sufficiently large
  critical amplitude to move the flow from the laminar into the
  congested phase. In stochastic models, it may be assumed that the
  size of this amplitude gets translated into a waiting time, i.e.\
  until fluctuations sufficiently add up to trigger the transition.  A
  recently introduced model of traffic flow however does not show this
  behavior: in the density regime where the jam solution co-exists
  with the high-flow state, the intrinsic stochasticity of the model
  is not sufficient to cause a transition into the jammed regime, at
  least not within relevant time scales. In addition, models can be
  differentiated by the stability of the outflow interface.  We
  demonstrate that this additional criterion is not related to the
  stability of the flow.  The combination of these criteria makes it
  possible to characterize commonalities and differences between many
  existing models for traffic in a new way.
\end{abstract}

\section{Introduction}

\begin{figure}[t]
\centerline{%
\includegraphics[width=0.8\hsize]{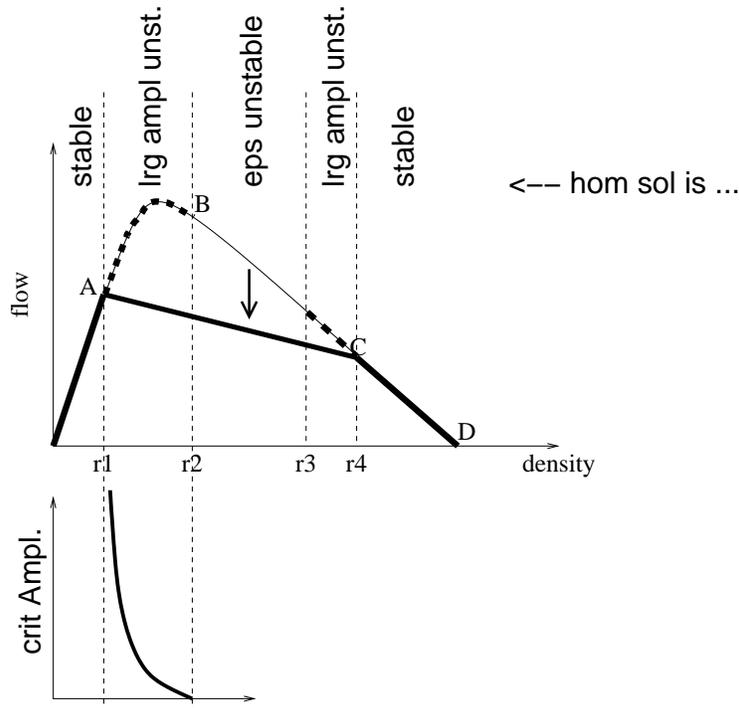}
}
\caption{Schematic fundamental diagram for deterministic models.
``lrg ampl unst.'' means that the homogeneous solution can be kicked
into an inhomogeneous state by a large enough amplitude; the bottom
plot gives a schematic graph for the necessary size of that critical
amplitude.  ``eps unstable'' means that the homogeneous solution is
linearly unstable.}
\label{fig:det}
\end{figure}

\begin{figure}[t]
\centerline{%
\includegraphics[width=0.8\hsize]{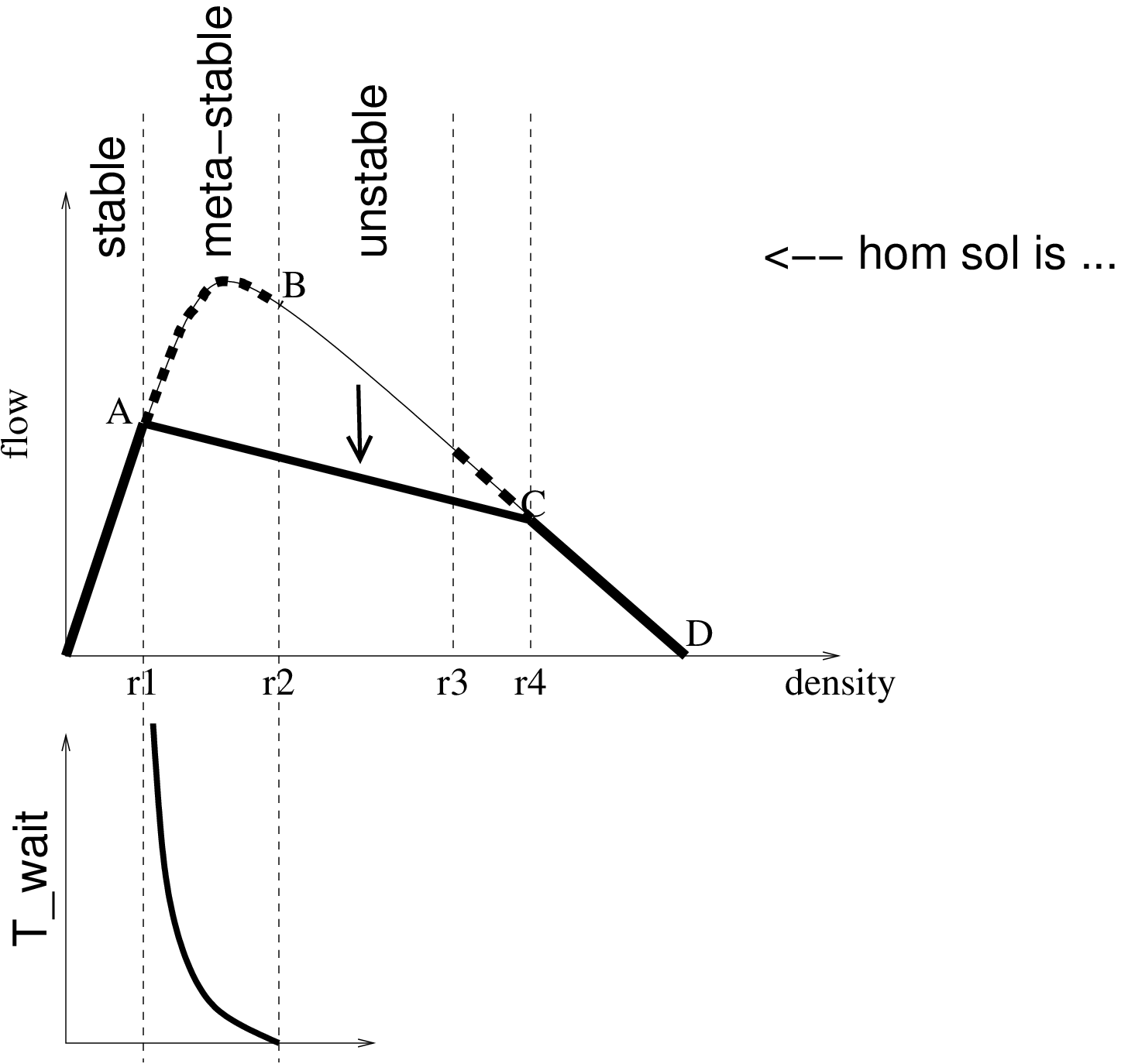}
}
\caption{Schematic fundamental diagram for stochastic models.
``meta-stable'' means that the homogeneous solution will be kicked
into an inhomogeneous state after a certain waiting time; the bottom
plot gives a schematic graph for that waiting time.}
\label{fig:stylized-noise}
\end{figure}

Car traffic is not always homogeneous.  For example, stop-and-go waves
are a frequently observed phenomenon.  Correspondingly, most traffic
models show a transition from laminar to ``congested'' flow and vice
versa.  For many deterministic models, this mechanism is well
understood (e.g.~\cite{Kerner:Konh:large:amplitude,Bando:etc:pre}, see
Fig.~\ref{fig:det}): For certain densities, the homogeneous solution
is linearly unstable, meaning that any tiny disturbance will destroy
the homogeneity and lead to another state, typically to one or more
waves.  For other densities, the homogeneous state may be linearly
stable, but unstable against large amplitude disturbances.

In stochastic models, one would intuitively assume (see
Fig.~\ref{fig:stylized-noise} that linear instability gets translated
into plain instability -- meaning that, for the corresponding densities,
the homogeneous state breaks down immediately -- and that for large
amplitude instability the large amplitude instability gets translated
into meta-stability -- meaning that, for the corresponding densities,
one has to wait some time until the noise conspires in a way that a
critical disturbance is generated and the instability is triggered.
This is exactly the topic for this paper, where we will demonstrate
that this speculation is correct in some cases but not in others.

Recent field measurements identify additional dynamic phenomena, such
as oscillations and so-called synchronized
traffic~\cite{Kerner:Rehborn:mea2,Helbing:review}.  It is under
discussion in how far these additional phenomena can be explained by
the above model instabilities in conjunction with geometrical
constraints (such as bottlenecks)~\cite{Lee:prl,Helbing:prl}, or if
additional features are necessary in the
models~\cite{Kerner:tgf01-talk}.  Given this, it seems desirable to
understand as much as we can about existing models.

Indeed, Krauss~\cite{Krauss:thesis} introduces model classes which he
names type~I, type~II, and type~III.  Type~III refers to a viscous,
syrup-like behavior without breakdown, and is not of relevance here.
Type~II displays jam formation, but jams have a typical size, meaning
that the system is macroscopically homogeneous, and that there is no
true phase transition.  Type~I displays true, macroscopic structure
formation and therefore a first order phase transition. In this paper,
we will argue that the Krauss characterization is incomplete.  We will
demonstrate that models can be stable or unstable at maximum flow, and
that the jams can have a stable or unstable interface.  The difference
to Krauss in Ref.~\cite{Krauss:thesis} is that he implicetely assumes
that stable maximum flow goes together with a stable interface, and
that unstable maximum flow goes together with an unstable interface.
Introducing our additional characterization means that we have $2
\times 2 = 4$ different classes, instead of just I and II.

In order to demonstrate this, we will first review what expectation
one has for traffic flow breakdown in analogy to a gas-liquid
transition~(Sec.~\ref{sec:gas-liquid}).  We then, in
Sec.~\ref{sec:model}, describe the traffic model that we use.  The
central Sections~\ref{sec:times} and~\ref{sec:ifaces} describe results
with respect to transition times, and with respect to interface
dynamics.  Sec.~\ref{sec:discussion} is a longer discussion of our
results, including speculations, conjectures, and some simulation
results for other models.  The paper is concluded by a summary.

\section{Traffic breakdown and the gas-liquid transition}
\label{sec:gas-liquid}

The breakdown of laminar traffic, i.e.\ the transition from
homogeneous traffic to stop-and-go waves, can be compared to a
gas-liquid transition, i.e.\ the transition from the homogeneous gas
state to the inhomogeneous gas/liquid coexistence state
(e.g.~\cite{Montroll,Krauss:thesis,dwolf:ca4traff:review}). As is well
known, if one compresses a gas beyond a certain critical density, then
it becomes super-critical, and small fluctuations will lead to droplet
formation and thus into the coexistence
state~\cite{Lifshitz:etc:book}. Similarly, we would expect for
homogeneous traffic that, once compressed beyond a certain critical
density, small fluctuations will lead to jam formation and thus into
the coexistence state.

And conversely, one knows that all droplets vanish once the mixture
is expanded beyond the critical density.  Similarly, one would
expect that all traffic jams vanish once the system is expanded
beyond a critical density.

This leads to predictions about the statistics of jam formation and
jam dissolution.  
Krauss~\cite{Krauss:thesis} gives the
following for the probability in a given time step that a jam starts
somewhere in the system:
\[
p_\downarrow \sim L \, \exp\left(\frac{\hat\alpha}{\mede{g} - g_0} \right) \ ,
\]
where $\mede{g} = 1/\rho = L/N$ is the inverse average density, $g_0$
and $\hat\alpha$ are free parameters, and $L$ is the system size.

The above is a bulk effect; jam formation can happen anywhere in the
system.  Jam dissolution, in contrast, is an interface effect: A jam
with $N$ vehicles dissolves if the random numbers come out the correct
way to let all $N_{jam}$ vehicles make the ``correct'' type of
movement.  This leads to a recovery probability of $p_\uparrow \sim
\exp(-N_{jam})$.  Since $N_{jam} \sim L \, (\rho - \rho_c)$, we obtain
\[
p_\uparrow \sim e^{- L \, (\rho-\rho_c)} \ .
\]
Note that $p_\uparrow$ is larger than zero also for $\rho > \rho_c$,
that is, spontaneous recovery should be possible in super-critical
systems although it becomes exponentially improbable with increasing
system size.  

In both cases, the time to recovery would be the inverse of the above
probabilities.  When setting the two equations equal, one obtains the
condition for a system to fluctuate back and forth between the
homogeneous and the coexistence state.  This would occur above
$\rho_c$; however, for any given $\rho > \rho_c$, in the case of $L
\to \infty$, $p_\downarrow$ would go to infinity while $p_\uparrow$
would go to zero, meaning that above $\rho_c$ only the coexistence
state is stable in the limit of $L
\to \infty$.

\section{The model}
\label{sec:model}

The model to be used in the following was introduced by
Krauss~\cite{Krauss:thesis}.  The basic idea is that cars drive as
fast as possible, but avoid crashes. Therefore, they have to choose
their velocity $v \leq v_{\rm safe}$ which takes into account the
braking distance $d(v)$ of the following and the braking distance
$d(\tilde v)$ of the preceding car. That means that the velocity has
to fulfil $d(v) + v\tau \leq d (\tilde v) + g$.  Here, $g$ is the
space headway between the cars given by $g = \tilde x - x -
\ell$. The braking capabilities of the cars are the same for all
cars and are parameterized by the maximum deceleration $b$. $\tau$ is
uniformly set to one throughout this paper.  This safety condition can
be transformed into a set of update rules as follows:
\begin{eqnarray}
v_{\rm safe} & = & \tilde v_t + 2\,b \frac{g_t - \tilde v_t}{2 \, b + v_t + \tilde v_t} \\
v_{\rm des} & = & \min \{ v_t + a, v_{\rm safe}, v_{\rm max} \} \\
v_{t+1} & = & \max \{v_{\rm des} - a \epsilon \xi, 0 \} \\
\label{eq:krauss}
x_{t+1} & = & x_t + v_{t+1} \ .
\end{eqnarray}
with index $t$ counting integer time.  The parameter $a$ is the
maximum acceleration, the parameter $\epsilon$ measures the degree of
randomness, $\xi$ is a random number, $\xi \in [0,1]$, while $v_{\rm
max}$ is the maximum velocity. We will use $v_{max}=3$ throughout this
paper.  Ref.~\cite{Krauss:thesis} discusses what our selection of
parameters means in terms of real world units; let us state that our
specific values have a reasonably close relation to the real world.

\section{Transition times}
\label{sec:times}

\begin{figure}[t]
\centerline{%
\includegraphics[width=0.8\hsize]{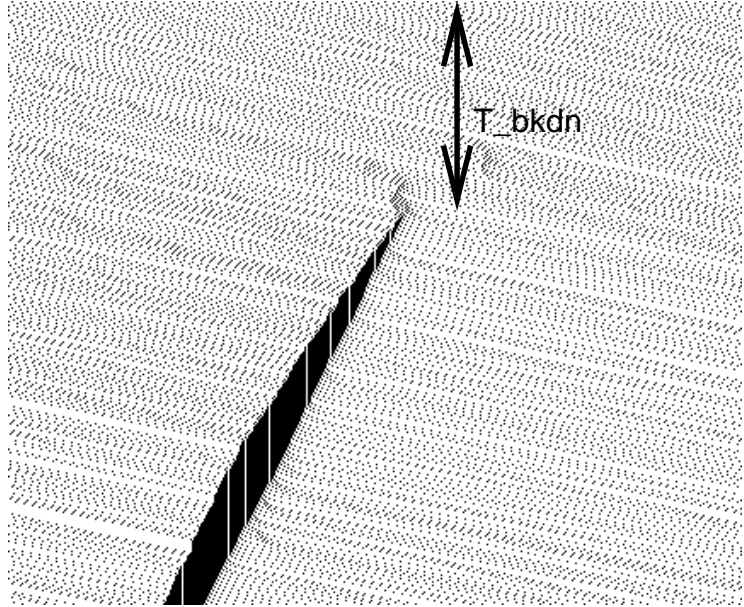}
}
\caption{Space-time plot for the breakdown time measurement.  Space is
horizontal; time increases downward; each line is a snapshot; vehicles
move from left to right.  Initially, all vehicles are lined up
equidistant with the specified density.  Time is measured until one
vehicle in the simulation comes to a complete stop ($v_i = 0$).  Once
a jam is started, it typically keeps growing until inflow is reduced,
either by another jam upstream, or by the effect of periodic boundary
conditions.}
\label{fig:bkdn-time}
\end{figure}

\begin{figure}[htbp]
\centerline{%
\includegraphics[width=0.49\hsize]{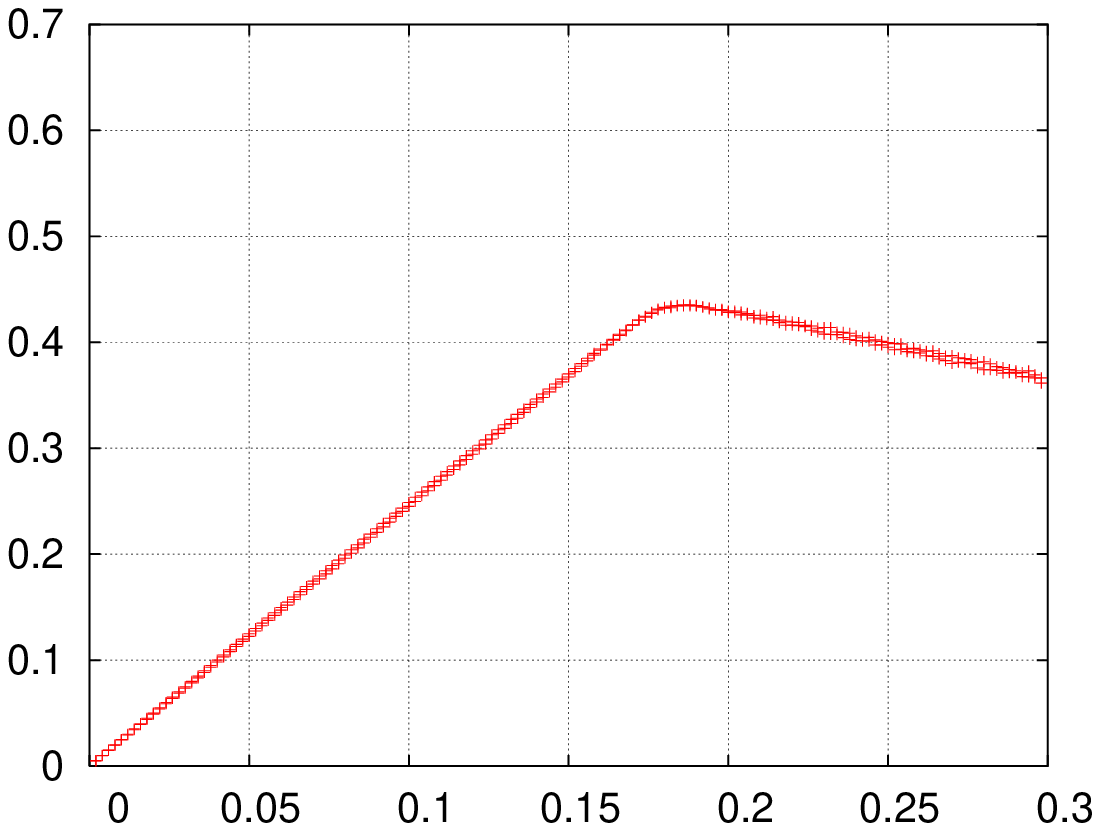}
\hfill
\hfill
\includegraphics[width=0.49\hsize]{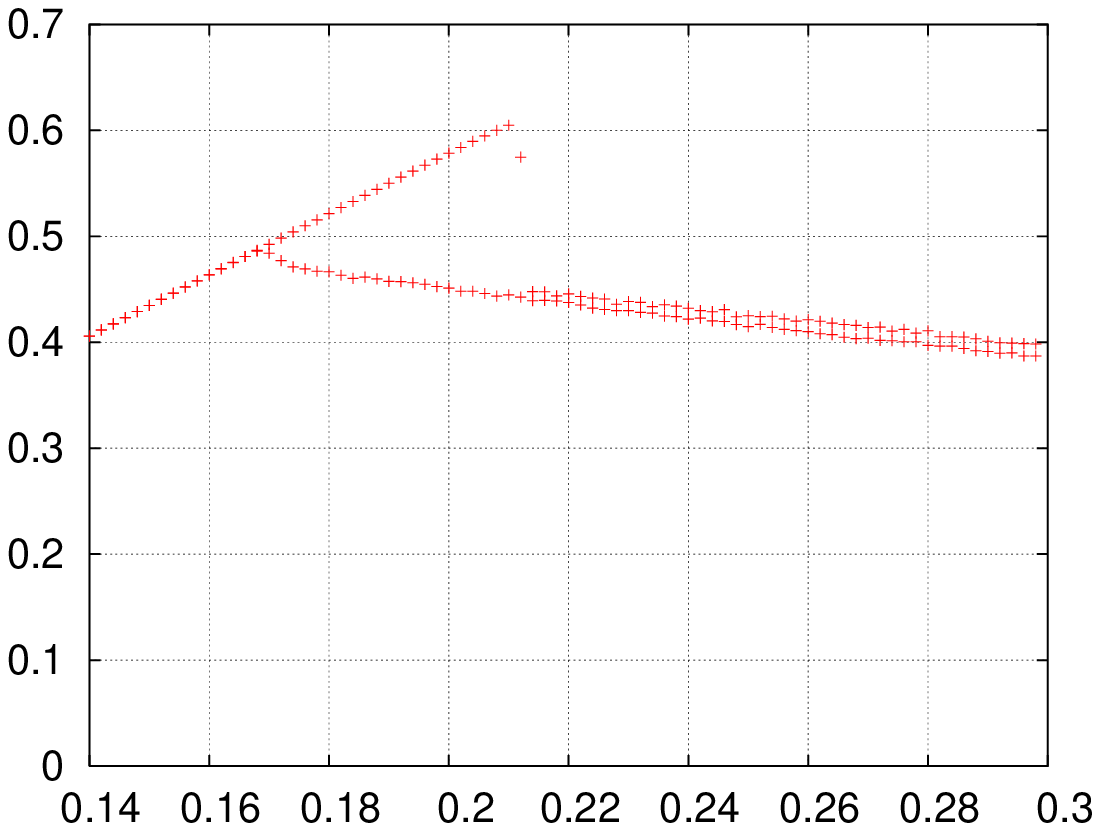}
}    

\centerline{%
\includegraphics[width=0.49\hsize]{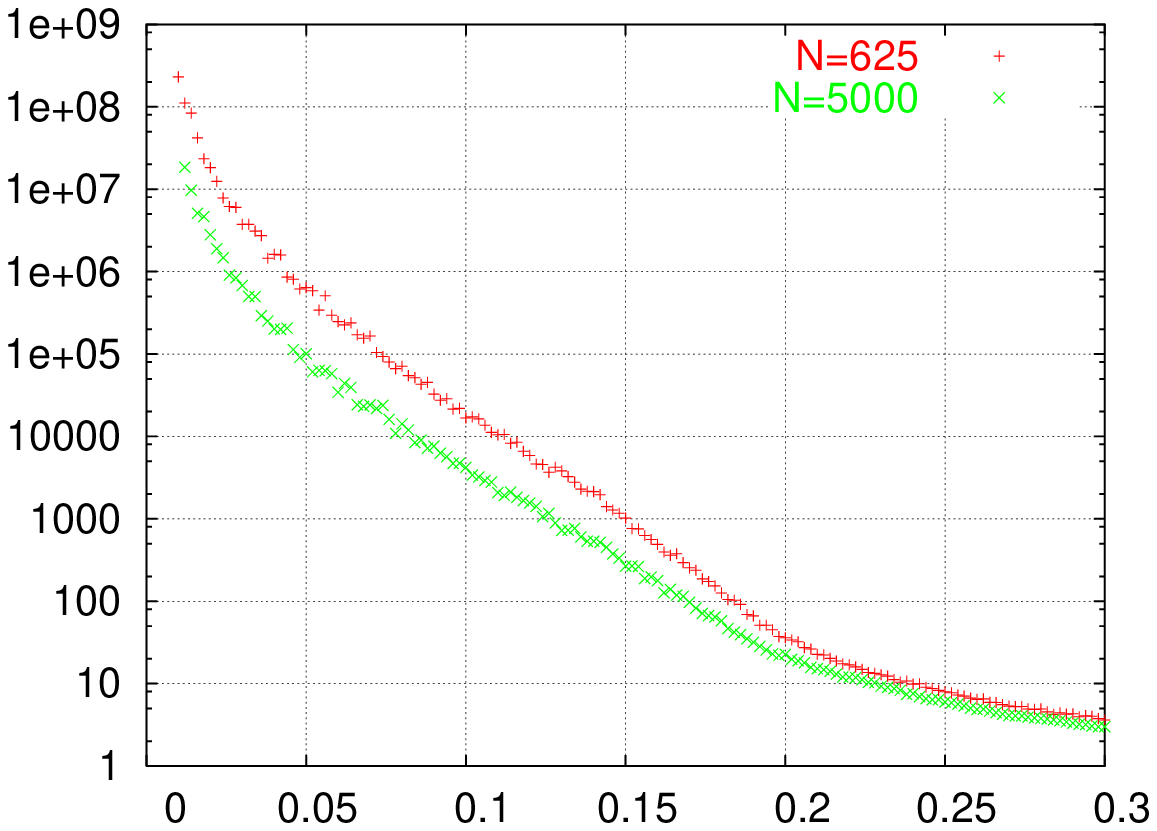}
\hfill
\hfill
\includegraphics[width=0.49\hsize]{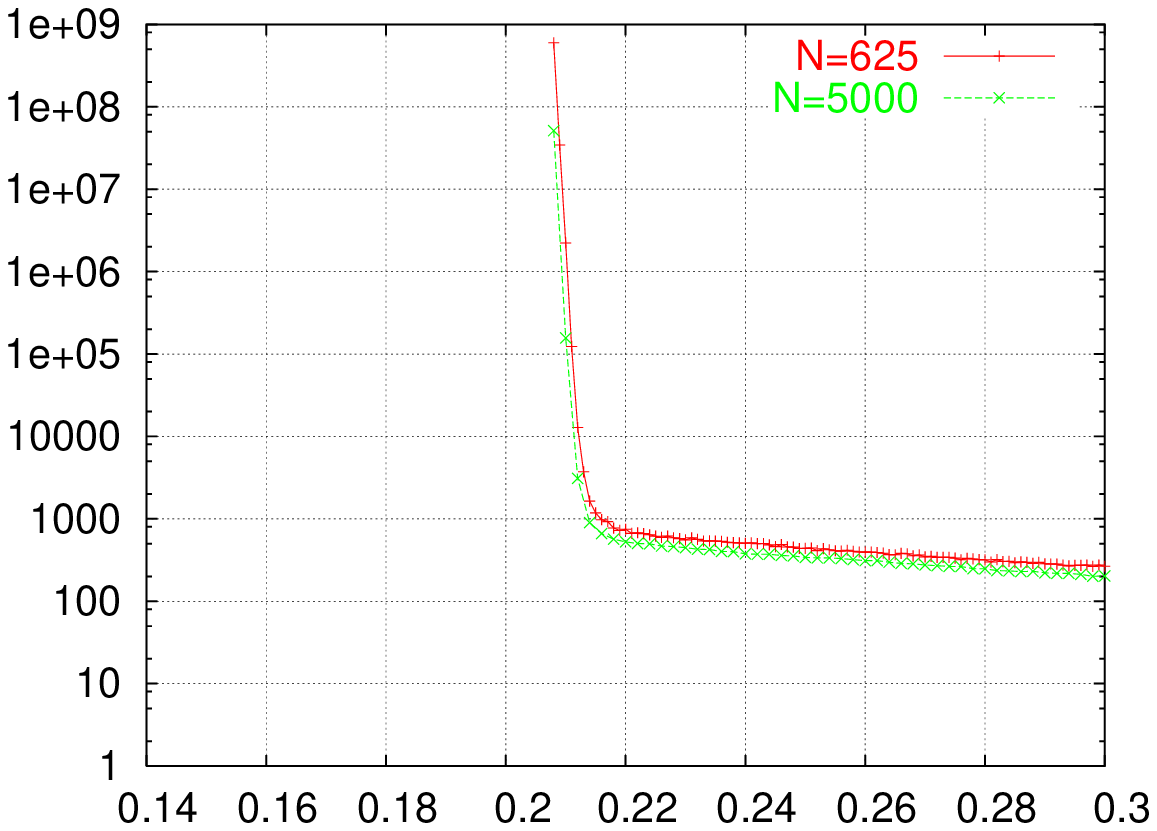}
}    
\centerline{%
\includegraphics[width=0.49\hsize]{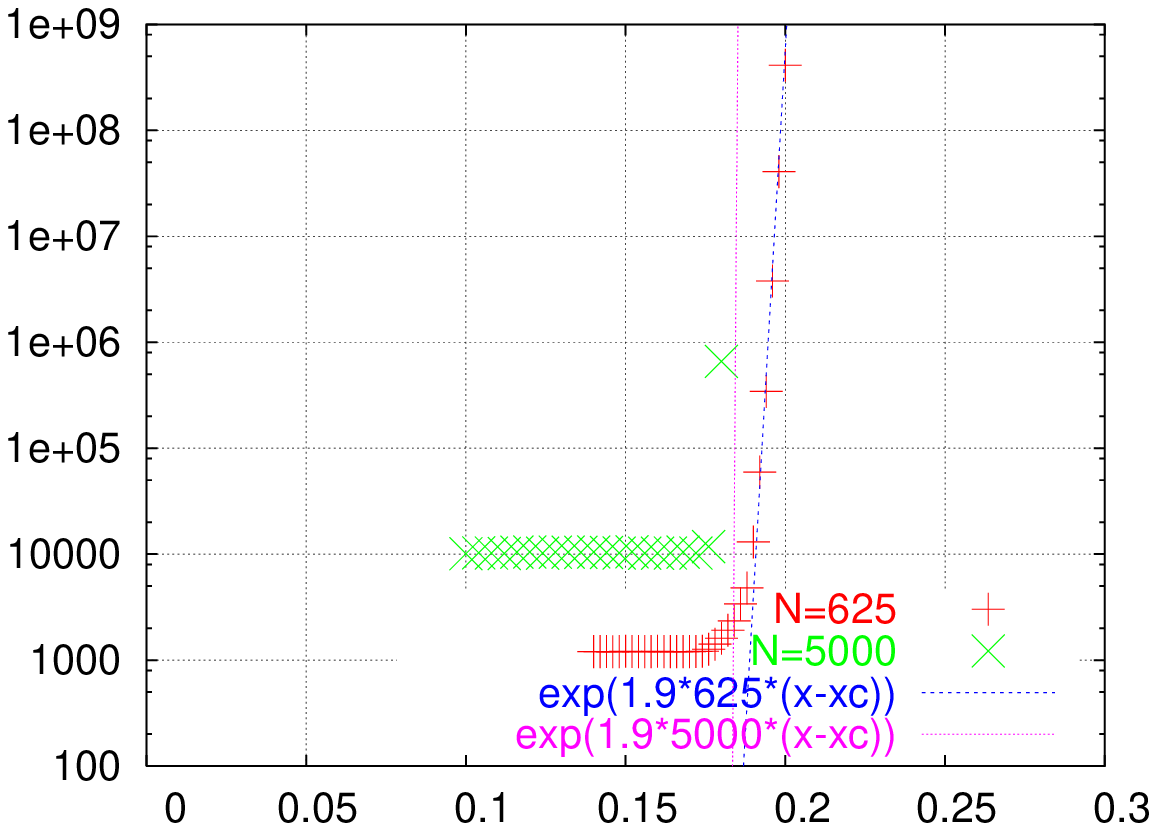}
\hfill
\hfill
\includegraphics[width=0.49\hsize]{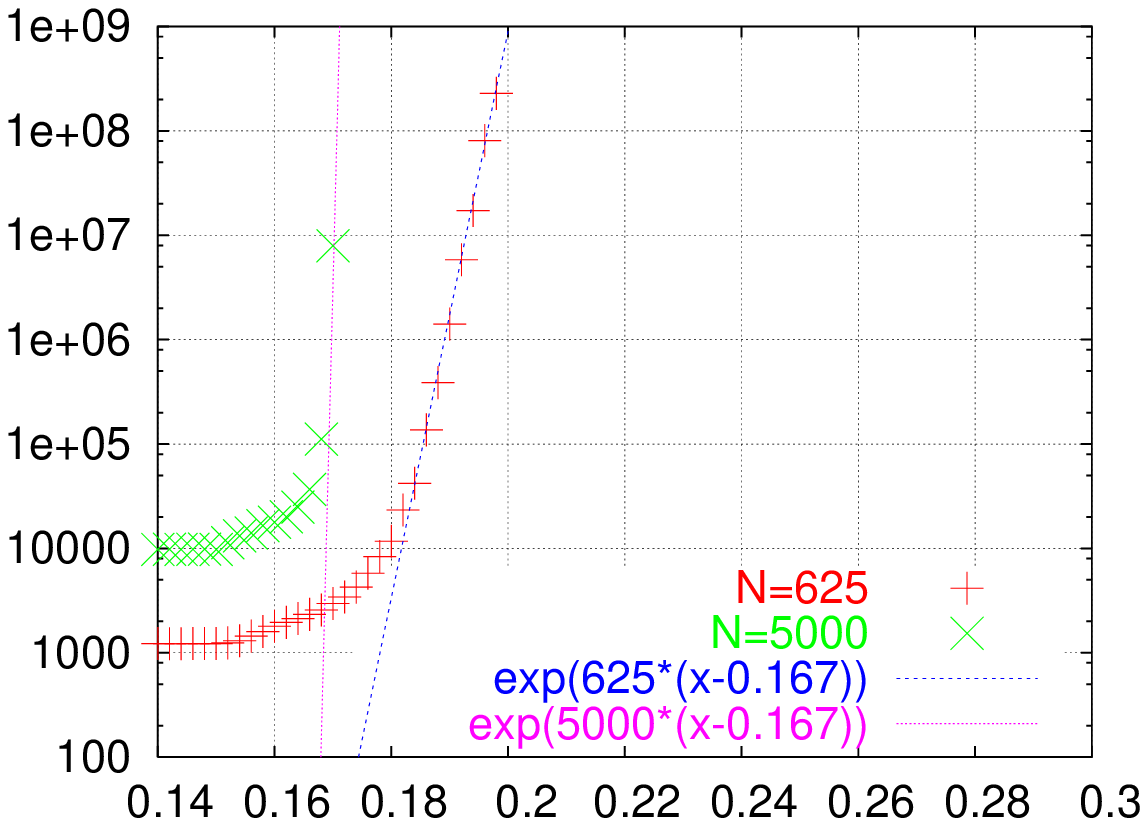}
}    
\label{fig:times-I}
\caption{%
Breakdown (middle) and recovery (bottom) times as function of density.
Left column: $(a,b) = (1,\infty)$.  Right column: $(a,b) = (0.2,0.6)$.
The straight lines in the bottom plots are proportional to
$\exp(A N \rho)$, where $N$ is the number of cars in the system, and
$A$ is a free parameter. -- In the right colum, we see that for
$N=5000$ there is a gap from $\rho \approx 0.17$ to $\rho \approx
0.205$, where the system is, up to $10^9$ time steps, stable both
against breakdown and against recovery. -- The corresponding sections
of the fundamental diagram (throughput as function of density; top)
for $N=625$ are given for orientation.  Each value of the fundamental
diagram is obtained at 5000~time steps; this is done once for
homogeneous and once for jammed initial condition, resulting in two
branches for bi-stable models.
}
\end{figure}

\begin{figure}[t]
\centerline{%
\includegraphics[width=0.8\hsize]{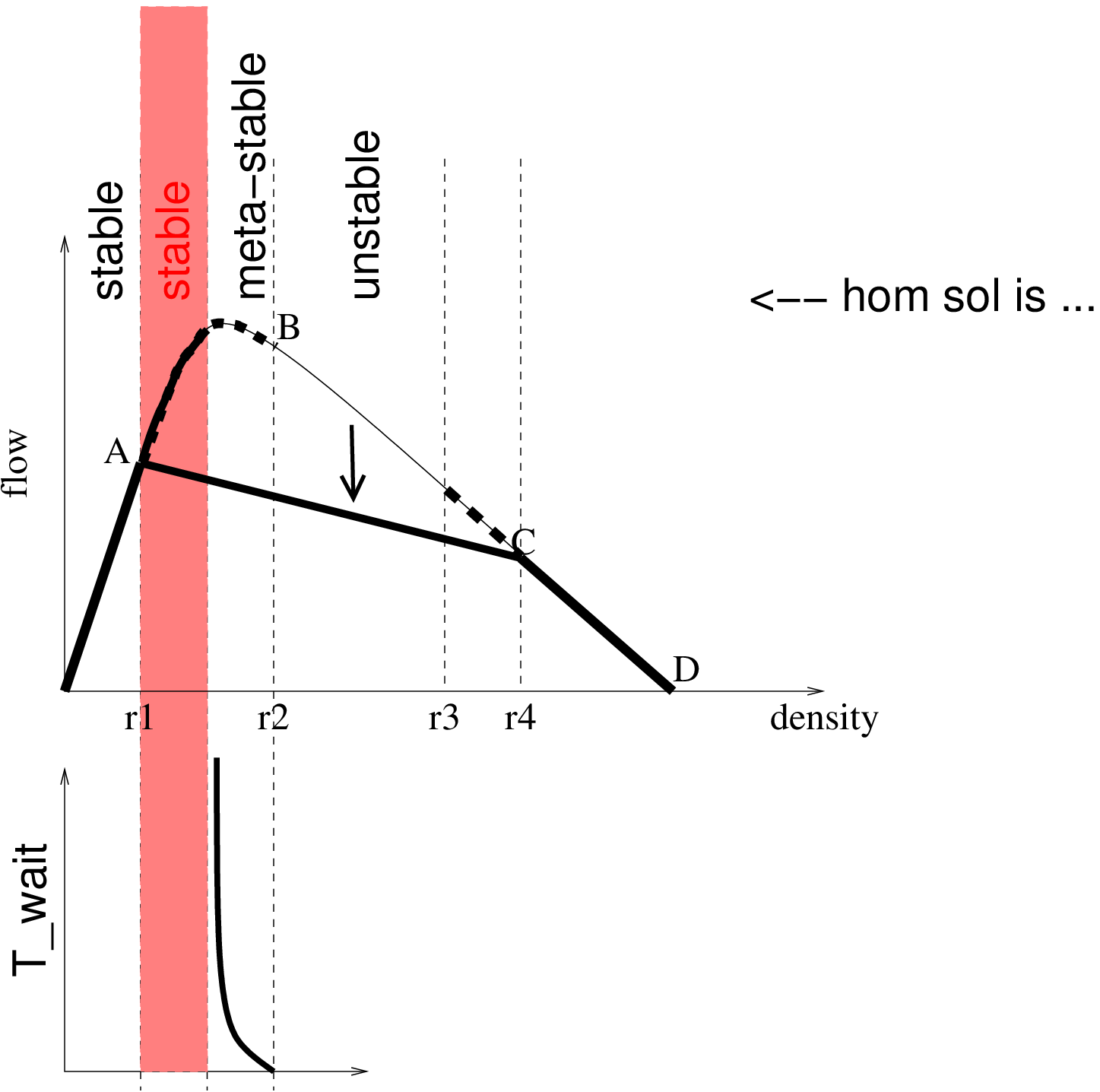}
}
\caption{Schematic fundamental diagram for stochastic models including
the new regime.  Compare to Fig.~\ref{fig:stylized-noise}.}
\label{fig:stylized-new}
\end{figure}

Fig.~\ref{fig:times-I} shows the breakdown and the recovery times for
two different sets of parameters: $(a,b) = (1,\infty)$ and 
$(a,b) = (0.2,0.6)$.  Recall that $a$ and $b$ are the acceleration and
braking capabilities, respectively.  The simulations are run with a
fixed number $N$ of vehicles; different densities are obtained by
adapting the system size $L$ via $L = N/\rho_L$.  The times are
obtained as follows:\begin{itemize}

\item \textbf{Breakdown times:} The system is started with all
  vehicles at equal distance $g = 1/\rho_L - 1$ and with velocity 
$v = \min \{g, v_{max} \}$ taken from the laminar branch of the
fundamental diagram.\footnote{%
Annoyingly, in the transition regime, for some parameters of $a$ and
$b$ different initial conditions lead to significantly different
breakdown times.  Outside the transition regime, the results are
robust.
} 
  The time is measured until the first vehicle in the system shows
  $v=0$ (see Fig.~\ref{fig:bkdn-time}).
\item \textbf{Recovery times:} The system is started with all
  vehicles except the leading one at distance one, i.e.\ gap equal to
  zero, and velocity zero.
  The time is measured until no vehicle with velocity zero is left
  in the system. 

\end{itemize}
Each data point is an average of at least 50~runs.
  
One observes from Fig.~\ref{fig:times-I} that, for both cases, the
recovery behavior is qualitatively consistent with the gas-liquid
transition picture: Above a certain $\rho_c$, the waiting time until
recovery (i.e.\ until a system with jams transitions to a system
without jams) shows exponential growth, which increases with system
size.

Similarly, for $(a,b)=(1,\infty)$, the breakdown results are
qualitatively consistent with the gas-liquid transition picture: The
time to breakdown \emph{de}creases with increasing system size, and it
increases with increasing density.  Putting breakdown and recovery
together, one obtains that for $L \to \infty$ and in equilibrium, a
system with $\rho > \rho_c$ should always be in the coexistence state.
For smaller $L$, the system can jump back and forth between
coexistence and the homogeneous state.

For parameters $(a,b)=(0.2,0.6)$, a possibly different picture
emerges.  Here, the breakdown times seem to diverge at $\rho^* \approx
0.2$, meaning that, for large $L$ and possibly for $L \to \infty$, we
have a density range where besides the transition from coexistence to
homogeneous also the inverse transition from homogeneous to
coexistence is extremely improbable.  \emph{That is, we may have a
stable supercritical homogeneous phase under an update rule that
includes noise.}  -- It is however very difficult to get good
numerical results for such fast growth as we find here: from
$\rho=0.22$ to $\rho=0.21$, the breakdown times grow between 2 and 5
orders of magnitude, depending on the system size.  Several function
fits were tried out, without convincing success; for example
$T \sim \exp\big( (\rho-\rho_*)^{-\alpha} \big)$ (implying divergence)
or
$T \sim \exp\big( - \gamma \, (\rho - \rho_*) \big)$ (implying no
divergence).  We cannot rule out the second functional form; however,
note that from Fig.~\ref{fig:times-I} we have, at $N=5000$, a gap
from $\rho\approx0.17$ up to $\rho \approx 0.21$ where both branches
are stable within times of $10^9$. -- Fig.~\ref{fig:stylized-new}
shows a schematic fundamental diagram with the new region.

Also, we note that breakdown times in general do not follow well the
analytical expectations.  Although qualitatively the dependency is as
expected (breakdown time decreasing with increasing system size), the
quantitative behavior is different.  The breakdown times for a given
density but different system sizes do not lie on a straight line on a
log-log plot (not shown), meaning that neither $T_{bkdn} \sim 1/L$ nor
any other algebraic form is applicable.  In contrast, for recovery
times, the simulation results are at least not inconsistent with
$T_{rcov} \sim \exp\big( - L(\rho - \rho_*) \big)$.  One should in
general note that very few system sizes were simulated; in particular,
the simulations do not extend to --computationally difficult-- very
large systems sizes where different behavior might be found.

Why is the Krauss model with $(a,b, \epsilon) = (0.2,0.6,1)$ so much
different from the ``standard'' phase transition picture, where noise
will relatively quickly add up in a way that a super-critical
homogeneous state will break down?  We suspect that in many traffic
models, because of the parallel update, noise is introduced in a
special way.  In particular, the amount of noise per spatial and
temporal unit is bounded.  In conjunction with a dynamics which
dissipates noise fast enough, it make sense to obtain states which are
absolutely stable under this kind of noise.  It is unclear to us if
the continuous variables used in the model here are a necessary
ingredient or not; preliminary simulation results indicate that the
same type of behavior can be obtained by a discrete model, but see
Ref.~\cite{Zielen:Schadschneider:arap} for a similar model where the
continuousness of the variables seems to play a crucial role.

\section{Interface dynamics}
\label{sec:ifaces}

\begin{figure}[t]
\centerline{%
\hfill
\includegraphics[height=0.4\textheight,width=0.4\hsize]{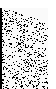}
\hfill
\includegraphics[height=0.4\textheight,width=0.4\hsize]{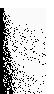}
\hfill
}
\caption{Space-time plot of interface dynamics.  As before, space is
pointing go the right and time is pointing down.  Space coordinates
are translated such that the leftmost edge of the moving traffic is
always at the same position.  LEFT: Krauss with
$(a,b,\epsilon) = (1,\infty,1)$.  Example for stable interface.
RIGHT: Krauss with $(a,b,\epsilon)=(0.2,0.6,1.5)$.  Example for
unstable interface. -- Note that for the left example, $a$ and $b$ are
selected in the range typically considered unstable, while for the
right example, $a$ and $b$ are selected in the range typically
considered stable.}
\label{fig:ifaces-pixels}
\end{figure}

\begin{figure}[t]
\centerline{%
\includegraphics[width=0.8\hsize]{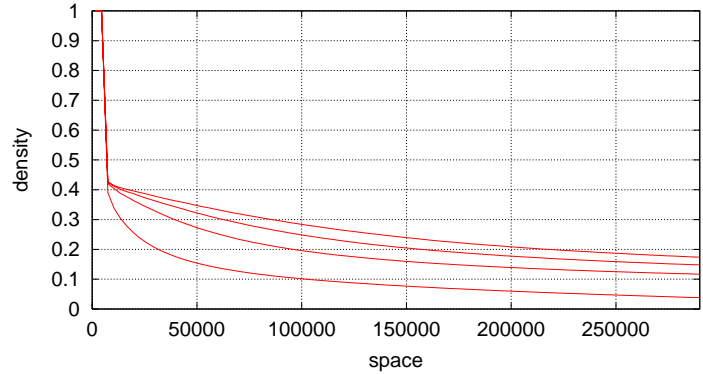}
}
\caption{Density profiles for $(a,b,\epsilon)=(1,\infty,1.5)$ at times
250\,000, 500\,000, 750\,000, and 1\,000\,000.  Clearly, the interface
keeps growing with time.  For the system on the left in
Fig.~\ref{fig:ifaces-pixels}, we obtain a completely stationary
interface profile, which also does not extend far into the system (not
shown).}
\label{fig:iface-krauss}
\end{figure}

\begin{table}[t]\footnotesize
\begin{center}
\begin{tabular}{|r|c|c|}
\hline
& stable outflow & unstable outflow \\
\hline
stable & E.g.\ $(a,b,\epsilon)=(0.2,0.6,1)$ & E.g.\ $(a,b,\epsilon)=(1,\infty,1)$ \\
i-face & ``Krauss type~I'' & \\
\hline
unstable & E.g.\ STCA cruice control & E.g.\ 
$(a,b,\epsilon) = (0.2,0.6,1.5)$\\
i-face & (Ref.~\protect\cite{Nagel:Paczuski}) & \\
\hline
\end{tabular}
\end{center}
\caption{Four cases of traffic breakdown.}
\label{table}
\end{table}

The nature of the transition (e.g.\ crossover vs.\ true phase
transition) is however not given by the time it takes until the first
fluctuation happens, but by how this fluctuation develops further, in
particular, if it spreads into the rest of the system or not.  In
order to further understand the nature of the transition, we will now
look at the dynamics of the interface between jam and outflow.  That
is, we start with an infinitely large mega-jam with $g=0$ and thus
$\rho=1$ in the half space from $x=-\infty$ to zero.  We collect data
for the development of the density profile as a function of time.
While doing that, we translate the zero of the coordinate system
always to the leftmost moving car, i.e.\ to the rightmost car in the
mega-jam which has not moved so far.  To the left from this point,
density is always one; in consequence, we look at the question if the
interface to the right will grow in time or if it will develop a
characteristic, time-independent profile.

Fig.~\ref{fig:ifaces-pixels} contains space-time plots of this
interface for two different systems.  In the left plot, the interface
is stable, whereas in the right plot, it keeps growing throughout the
plot.  The left plot is obtained with $(a,b,\epsilon)=(1,\infty,1)$,
which is one of the two models for which we have investigated the
transition times in more detail above.  A plot for
$(a,b,\epsilon)=(0.2,0.6,1)$ looks similar (not shown).  In contrast,
the plot on the right with the growing interface is obtained with a
larger noise amplitude, i.e.~$(a,b,\epsilon)=(0.2,0.6,1.5)$.

In order to investigate the long-term behavior, we also plotted
density profiles at different times.  A stable interface is
characterized by a density profile which eventually becomes
stationary; an unstable interface keeps growing.
Fig.~\ref{fig:iface-krauss} contains a result for the model of
Fig.~\ref{fig:ifaces-pixels} right, i.e.\ with 
$(a,b,\epsilon) = (0.2,0.6,1.5)$.  The plot contains density profiles
at times 250\,000, 500\,000, 750\,000, and 1\,000\,000.  Each curve is
the average of 60~runs.  Clearly, the plot shows that the interface
grows with time.  In fact, when looking at a fixed density value, say
$\rho=0.2$, it seems that the interface width is growing linearly in
time.

What this means is that \emph{the stability of the interface is a
property which is separate from the stability of the flow}.
Table~\ref{table} lays out the resulting four cases.  The finding of
$2 \times 2$ criteria goes beyond the findings of
Krauss~\cite{Krauss:thesis}, who only differentiates between stable
(``Type~I'') and unstable (``Type~II'') maximum flow.  Krauss mentions
``branching'', but more or less explicitely assumes that branching
goes along with unstable maximum flow.  In addition, the example of
Ref.~\cite{Krauss:thesis} for ``branching''
($(a,b,\epsilon)=(1,\infty,1)$) in fact has a stable interface as
demonstrated in Fig.~\ref{fig:ifaces-pixels}.

\section{Discussion and open questions}
\label{sec:discussion}

\begin{figure}[t]
\centerline{%
\includegraphics[width=0.8\hsize]{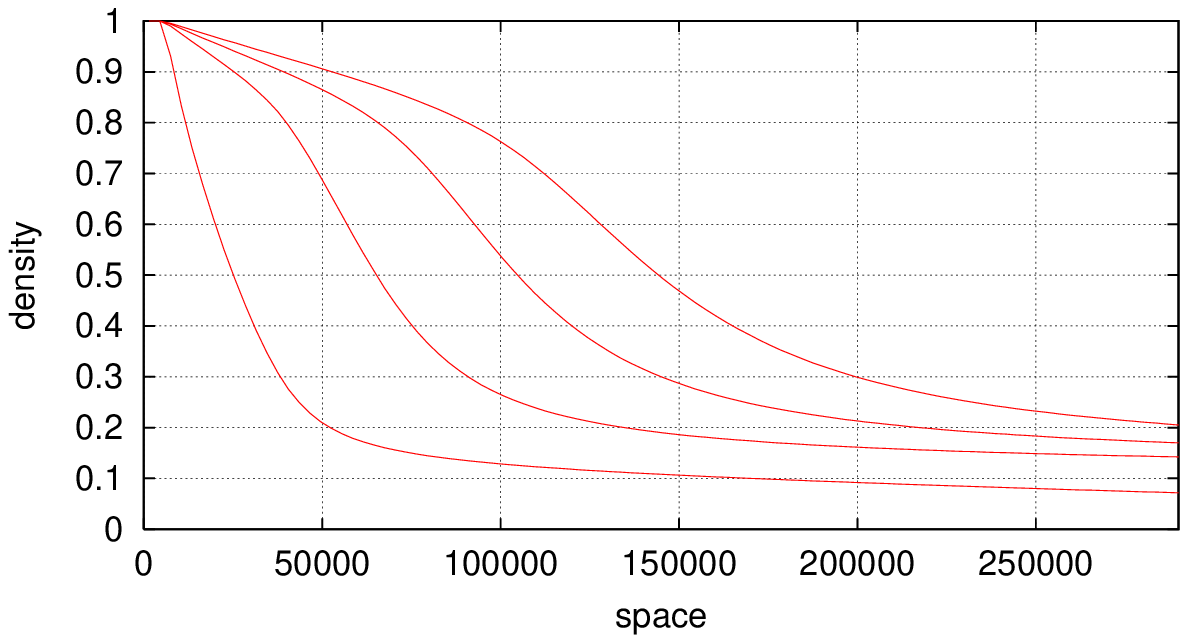}
}
\centerline{%
\includegraphics[width=0.8\hsize]{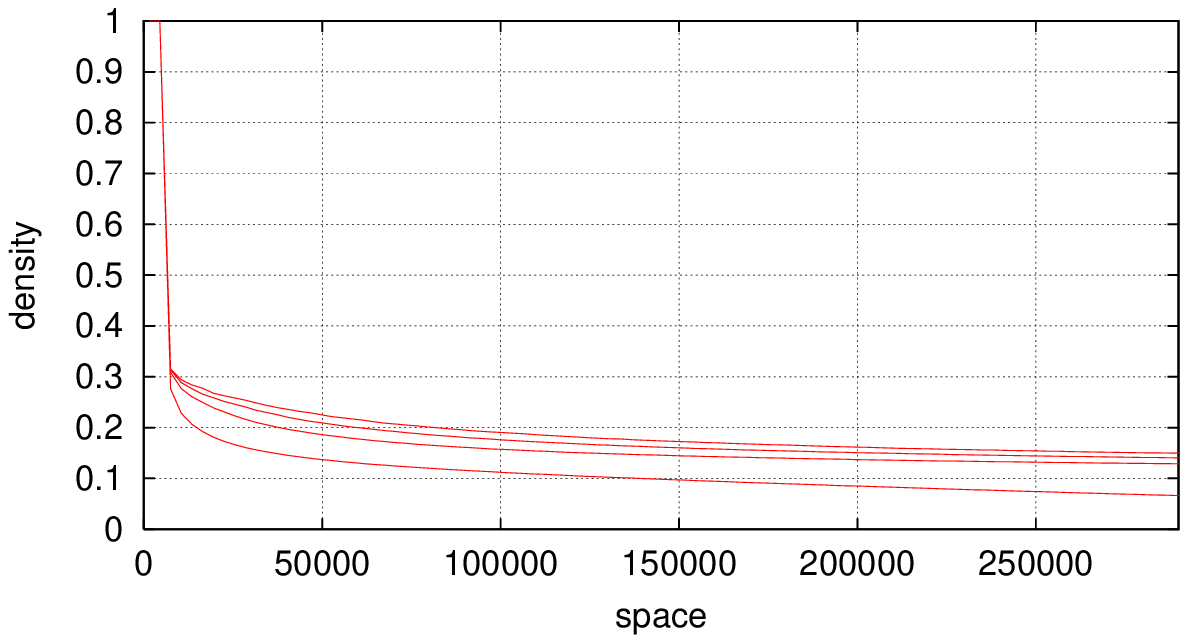}
}
\caption{Density profiles at times
250\,000, 500\,000, 750\,000, and 1\,000\,000 for the stochastic
traffic cellular automaton (STCA) of Ref.~\cite{Nagel:Schreck} (top) and for
the slow-to-start model of Ref.~\cite{Barlovic} (bottom).  Clearly, in
both cases the interface is non-stationary.}
\label{fig:ifaces-ca}
\end{figure}

There is controversy if cellular automata (CA) models for traffic
show a first order phase transition,\footnote{%
  In Ref.~\cite{dwolf:ca4traff:review}, Wolf reviews evidence for
  and against true phase coexistence, without making a judgement.
  } a true critical phase
  transition~\cite{Roters:etc:critical:traffic}, or none at
  all~\cite{Sasvari:Kertesz,Chowdhury:etc:comment}.  The discussion
  was seriously hampered by the fact that no parameter was known to
  change the possibly critical behavior of the system.  Our findings,
  with a different type of model, shed new light on this discussion.
  It is plausible to assume that models with an unstable outflow
  interface display a crossover behavior, because any phase separation
  in the initial conditions will spread through the system -- in a
  finite system, there would eventually be a macroscopically
  homogeneous state although there would be structure on the
  microscopic scale.  Conversely, models with a stable outflow
  interface will display true macroscopic phases.  Since the different
  phases are obtained by variations of continuous parameters, it
  should be possible (albeit computationally expensive) to find the
  line in phase space which separates the two regimes.

Additional simulations show that the stochastic traffic cellular
automaton (STCA) of Ref.~\cite{Nagel:Schreckenberg} has indeed both an
unstable outflow and an unstable interface (Fig.~\ref{fig:ifaces-ca}
top).  The so-called cruise control limit of this
model~\cite{Nagel:Paczuski} also has an unstable interface but a
stable outflow, although marginally so.  We also tested the so-called
slow-to-start model~\cite{Barlovic} and found that it has, for the
parameters that we tested, an unstable interface
(Fig.~\ref{fig:ifaces-ca} bottom).  This puts it in a class separate
from Krauss type~I, in contrast to the original motivation that it
would display the same kind of ``meta''-stability as a Krauss type~I
model.

Wolf, in Ref.~\cite{dwolf:ca4traff:review}, describes a so-called
Galileo-invariant CA traffic model, where he observes a different type
of meta-stability than the slow-to-start models.  It is open into
which of our four classes that model belongs.

In summary, it seems that our findings are finally the starting point
of a more complete classification of the different models for traffic.
Also from an engineering/applications perspective, it is necessary to
solve these questions because of their consequences for real world
applications.  For example, the existence of stable high-flow states
under noise would mean that it should be possible to stabilize these
states in the real world.  And an unstable outflow interface would
imply different interpretations of real world data, which are
typically averages over 1~minute or longer.

An implication of our findings is that, in contrast to earlier claims,
outflow is not constant in STCA-type models: It is constant only
outside the boundary region, which however grows to infinity.  On the
other hand, for both Krauss models of this paper, with 
$(a,b) = (1,\infty)$ and $(a,b) = (0.2,0.6)$, and noise $\epsilon$
small enough, outflow is indeed a constant.

This implies that our theory about breakdown behavior in microscopic
models needs to be revised.  That theory was that there is a
characteristic jam outflow, and any homogeneous solution with higher
densities would be unstable against large amplitude disturbances, such
as stopping a vehicle and releasing it later.  For models where the
outflow is not well defined this is obviously too simplistic.

In addition, it seems that also for models with stable interfaces the
situation can be more complicated.  Our own simulations show that,
essentially, the density between jams can be ``compressed'' in models
with continuous variables.  This is not discussed further here.

\section{Summary}
\label{sec:summary}

We have demonstrated that the breakdown of the homogeneous state in
stochastic traffic models is characterized by two properties:
(i)~stability or not of the high flow states; (ii)~stability or not of
the outflow interface of jams.  This is different from earlier
findings, where it was assumed that the two go together.  This is
important, since it will allow to characterize the different existing
traffic models according to these properties.  It should also allow to
eventually settle the controversy over the nature of the transition
from homogeneous to congested flow.  Engineering applications should
benefit from these findings by being able to pick the model type which
closest reflects reality.  And finally, it is an interesting physical
question since we are looking at simple one-dimensional driven systems
which display interesting dynamics and which can be analyzed using the
methods of statistical physics.

\section{Acknowledgements}

We had the suspicion that the Krauss model behaved differently from
the slow-to-start rules of CA models for some time. This suspicion was
reinforced by discussions with D.~Helbing and with D.~Wolf, who
reported similar observations from some of their models.
Additionally, simulations performed by N.~Eissfeldt helped to clarify
some of the questions related to this different behaviour.  Computing
time of about 100\,000~CPU-hours on the clusters Asgard and Xibalba,
both at ETHZ, is acknowledged.

\bibliographystyle{unsrt}
\bibliography{ref,kai,xref}

\begin{thebibliography}{10}

\bibitem{Kerner:Konh:large:amplitude}
B.~S. Kerner and P.~Konh\"auser.
\newblock Structure and parameters of clusters in traffic flow.
\newblock {\em Physical Review E}, 50(1):54, 1994.

\bibitem{Bando:etc:pre}
M.~Bando, K.~Hasebe, A.~Nakayama, A.~Shibata, and Y.~Sugiyama.
\newblock Dynamical model of traffic congestion and numerical simulation.
\newblock {\em Physical Review E}, 51(2):1035, 1995.

\bibitem{Kerner:Rehborn:mea2}
B.~S. Kerner and H.~Rehborn.
\newblock Experimental properties of complexity in traffic flow.
\newblock {\em Physical Review E}, 53(5):R4275--R4278, 1996.

\bibitem{Helbing:review}
D.~Helbing.
\newblock Traffic and related self-driven many-particle systems.
\newblock {\em Reviews of Modern Physics}, Oct/Nov 2001.
\newblock Also www.arXiv.org, cond-mat/0012229.

\bibitem{Lee:prl}
H.Y. Lee, H.-W. Lee, and D.~Kim.
\newblock Origin of synchronized traffic flow on highways and its dynamic phase
  transitions.
\newblock {\em Physical Review Letters}, 81(5):1130--1133, August 1998.

\bibitem{Helbing:prl}
D.~Helbing and M.~Treiber.
\newblock Gas-kinetic-based traffic model explaining observed hysteretic phase
  transition.
\newblock {\em Physical Review Letters}, 81(14):3042--3045, 1998.

\bibitem{Kerner:tgf01-talk}
B.~Kerner.
\newblock presented at Traffic and Granular Flow '01 in Nagoya, Japan.

\bibitem{Krauss:thesis}
S.~Krau{\ss}.
\newblock {\em Microscopic modeling of traffic flow: {I}nvestigation of
  collision free vehicle dynamics}.
\newblock PhD thesis, University of {C}ologne, Germany, 1997.
\newblock See www.zpr.uni-koeln.de.

\bibitem{Montroll}
E.W. Montroll.
\newblock Theory and observations of the dynamics and statistics of traffic on
  an open road, 1962.

\bibitem{dwolf:ca4traff:review}
D.~Wolf.
\newblock Cellular automata for traffic simulations.
\newblock {\em Physica A}, 263:438--451, 1999.

\bibitem{Lifshitz:etc:book}
E.M. Lifschitz and L.P. Pitajewski.
\newblock Statistische physik, {T}eil 1.
\newblock In L.D. Landau and E.M. Lifschitz, editors, {\em Statistische Physik,
  {T}eil 1}, volume~I of {\em Lehrbuch der Theoretischen Physik}.
  Akademie-Verlag, 1987.

\bibitem{Zielen:Schadschneider:arap}
F.~Zielen and A.~Schadschneider.
\newblock Nonsymmetric ergodicity breaking in a stochastic model on continuous
  phase space.
\newblock {\em preprint}, 2001.

\bibitem{Nagel:Paczuski}
K.~Nagel and M.~Paczuski.
\newblock Emergent traffic jams.
\newblock {\em Physical Review E}, 51:2909, 1995.

\bibitem{Nagel:Schreck}
K.~Nagel and M.~Schreckenberg.
\newblock A cellular automaton model for freeway traffic.
\newblock {\em Journal de Physique I France}, 2:2221, 1992.

\bibitem{Barlovic}
R.~Barlovic, L.~Santen, A.~Schadschneider, and M.~Schreckenberg.
\newblock Metastable states in cellular automata.
\newblock {\em European Physical Journal B}, 5(3):793--800, 10 1998.

\bibitem{Roters:etc:critical:traffic}
L.~Roters, S.~L\"ubeck, and K.D. Usadel.
\newblock Critical behavior of a traffic flow model.
\newblock {\em Physical Review E}, 59:2672, 1999.

\bibitem{Sasvari:Kertesz}
M.~Sasvari and J.~Kertesz.
\newblock Cellular automata models of single lane traffic.
\newblock {\em Physical Review E}, 56(4):4104--4110, 1997.

\bibitem{Chowdhury:etc:comment}
D.~{Chowdhury et al}.
\newblock Comment on: {``Critical behavior of a traffic flow model''}.
\newblock {\em Physical Review E}, 61(3):3270--3271, 2000.

\bibitem{Nagel:Schreckenberg}
K.~Nagel and M.~Schreckenberg.
\newblock A cellular automaton model for freeway traffic.
\newblock {\em Journal de Physique I France}, 2:2221, 1992.

\end{thebibliography}

\end{document}